\documentclass[a4paper,11pt]{article}
\pdfoutput=1 

\usepackage{jinstpub} 

\usepackage{color}
\usepackage{lineno}
\usepackage{comment}
\usepackage{siunitx}
\usepackage{orcidlink}
\usepackage{tikz}
\usetikzlibrary{circuits.logic.US}
\usepackage{siunitx}
\title{\boldmath Development of proton beam irradiation system for the NA65/DsTau experiment}

\newcommand{\Yuri}[1]{\textcolor{red}{[YG: {#1}]}}

\collaboration[]{The DsTau collaboration}
\author[a]{Shigeki Aoki\orcidlink{0000-0002-1092-5037},}
\author[b,c]{Akitaka Ariga\orcidlink{0000-0002-6832-2466},}
\author[d]{Tomoko Ariga\orcidlink{0000-0001-9880-3562},}
\author[e]{Nikolaos Charitonidis\orcidlink{0000-0001-9506-1022},}
\author[f]{Sergey Dmitrievsky\orcidlink{0000-0003-4247-8697},}
\author[g,l]{Radu Dobre\orcidlink{0000-0002-9518-6068},}
\author[g]{Elena Firu\orcidlink{0000-0002-3109-5378},}
\author[f]{Yury Gornushkin\orcidlink{0000-0003-3524-4032},}
\author[h]{Ali Murat Guler\orcidlink{0000-0001-5692-2694},}
\author[b]{Daiki Hayakawa\orcidlink{0000-0003-4253-4484},}
\author[i]{Koichi Kodama\orcidlink{0000-0001-9533-1571},}
\author[j]{Masahiro Komatsu\orcidlink{0000-0002-6423-707X},}
\author[k]{Umut Kose\orcidlink{0000-0001-5380-9354},}
\author[f,l]{M\u{a}d\u{a}lina-Mihaela Miloi\orcidlink{0000-0001-7208-4379},}
\author[b]{Manato Miura\orcidlink{0000-0002-4955-8609},}
\author[j]{Mitsuhiro Nakamura\orcidlink{0009-0002-6032-2741},}
\author[j]{Toshiyuki Nakano\orcidlink{0009-0004-8568-9077},}
\author[g]{Alina-Tania Neagu\orcidlink{0000-0001-6788-4320},}
\author[b,*]{Toranosuke Okumura\note[*]{Corresponding author.}\orcidlink{0000-0002-3266-8713},}
\author[h]{Canay Oz\orcidlink{0000-0002-5113-5779},}
\author[j]{Hiroki Rokujo\orcidlink{0000-0002-3502-493X},}
\author[j]{Osamu Sato\orcidlink{0000-0002-6307-7019},}
\author[f]{Svetlana Vasina\orcidlink{0000-0003-2775-5721},}
\author[m]{Junya Yoshida\orcidlink{0000-0002-9398-746X},}
\author[n]{Masahiro Yoshimoto\orcidlink{0000-0002-4667-0718},}
\author[h]{Emin Yuksel\orcidlink{0009-0008-7861-1879}.}

\affiliation[a]{Graduate School of Human Development and Environment, Kobe University, Tsurukabuto, Nada, 657-8501 Kobe, Japan}

\affiliation[b]{Department of Physics, Chiba University, 1-33 Yayoi-cho Inage-ku, 263-8522 Chiba, Japan
}

\affiliation[c]{Albert Einstein Center for Fundamental Physics, Laboratory for High Energy Physics, University of Bern, Sidlerstrasse 5, CH-3012 Bern, Switzerland}

\affiliation[d]{Faculty of Arts and Science, Kyushu University, 744 Motooka, Nishi-ku, Fukuoka, 819-0395 Japan}

\affiliation[e]{CERN, BE Department, 1 Esplanade des Particules, CH-1211 Meyrin, Switzerland}

\affiliation[f]{Affiliated with an international laboratory covered by a cooperation agreement with CERN}

\affiliation[g]{Laboratory of High Energy Astrophysics and Advanced Technology, Institute of Space Science a subsidiary of INFLPR, 409, Atomistilor Street, Magurele, 077125 Ilfov, Romania}

\affiliation[h]{Physics Department, Middle East Technical University, Dumlup{\i}nar Bulvari, 06800 Ankara, Turkey}

\affiliation[i]{Department of Science Education, Aichi University of Education, 448-8542 Kariya, Japan}

\affiliation[j]{Department of Physics, Nagoya University, Furo-cho, Chikusa-ku, 464-8602 Nagoya, Japan}

\affiliation[k]
{Institute for Particle physics and Astrophysics, ETH Zurich, Otto-Stern-Weg 5, CH-8093 Zurich, Switzerland}

\affiliation[l]
{Faculty of Physics, University of Bucharest, 077125 Bucharest, Romania}

\affiliation[m]{Tohoku University, Sendai city, 980-8577 Sendai, Japan}

\affiliation[n]{RIKEN Nishina Center, RIKEN, 2-1 Hirosawa, Wako, 351-0198 Saitama, Japan}

\emailAdd{toranosuke.okumura@cern.ch}

\abstract{
Tau neutrino is the least studied lepton of the Standard Model (SM).  
The NA65/DsTau experiment targets to investigate $D_s$, the parent particle of the $\nu_\tau$, 
using the nuclear emulsion-based detector and to decrease the systematic uncertainty of $\nu_\tau$ flux prediction from over $50\si{\%}$ to 10\si{\%} for future beam dump experiments. 
In the experiment, the emulsion detectors are exposed to the CERN SPS 400 GeV proton beam. 
To provide optimal conditions for the reconstruction of interactions, the protons are required to be uniformly distributed over the detector's surface with an average density of $10^5 \si{cm^{-2}}$ and the fluctuation of less than 10\%.
To address this issue, we developed a new proton irradiation system called the target mover.
The new target mover provided irradiation with a proton density of \SI{1.01e5}{cm^{-2}} and the density fluctuation of $1.9\pm0.3$\% in the DsTau 2021 run.
}


\begin{document}
\maketitle
\flushbottom

\section{Introduction}
\label{sec:intro}

The validation of the Standard Model (SM) and exploration of Beyond Standard Model (BSM) physics are considered to be a paramount mission in particle physics.
Recent results from the LHCb \cite{LHCb:2015gmp}, BaBar \cite{BaBar:2001pki}, and Belle \cite{Belle:2001zzw} (Sec. 7.6 in \cite{HFLAV:2022pwe}) demonstrate hints of possible violation of the Lepton Universality (LU) in $B$ meson decays. 
The study of LU in neutrino interactions can be a new probe for BSM. 
However, the data on $\nu_\tau$ is quite scarce; only a few experiments have reported its detection.
The DONuT experiment \cite{DONUT:2000fbd} directly detected $\nu_\tau$ for the first time and estimated the $\nu_\tau$ interaction cross-section \cite{Furukawa:2008zza}.
However, the cross-section measurement had about 30\% 
statistics error due to the low statistics and about 50\% systematic error due to a poorly constrained $\nu_\tau$ flux.
The main source of $\nu_\tau$ is the leptonic decay of $D_s$ mesons.
Therefore, a precise measurement of the $D_s$ production cross-section can provide prediction of $\nu_\tau$ fluxes for neutrino experiments like FASER($\nu$) \cite{FASER:2019dxq, FASER:2020gpr}, SND@LHC \cite{SNDLHC:2022ihg} and future experiments proposed at CERN BDF \cite{Aberle:2022fts}.
The NA65/DsTau experiment \cite{DsTau:2019wjb, Aoki:2017spj} at CERN-SPS was proposed to measure $D_s$ production cross-section in proton-nucleus interactions by detecting about $10^3$ $D_s\to\tau\to X$ decays.
This measurement is going to reduce uncertainty in the DONuT's measurement from 50\% to 10\%.


The identification of $D_s\to \tau \to X$ decays will be performed by using topological information, thanks to the high spatial and angular resolution of the emulsion-based detectors~\cite{Amsler_2013}.

\begin{figure}[tbp]
\centering
\includegraphics[width=.4\textwidth]{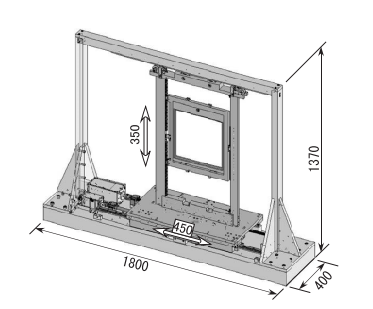}
\qquad
\includegraphics[width=.4\textwidth]{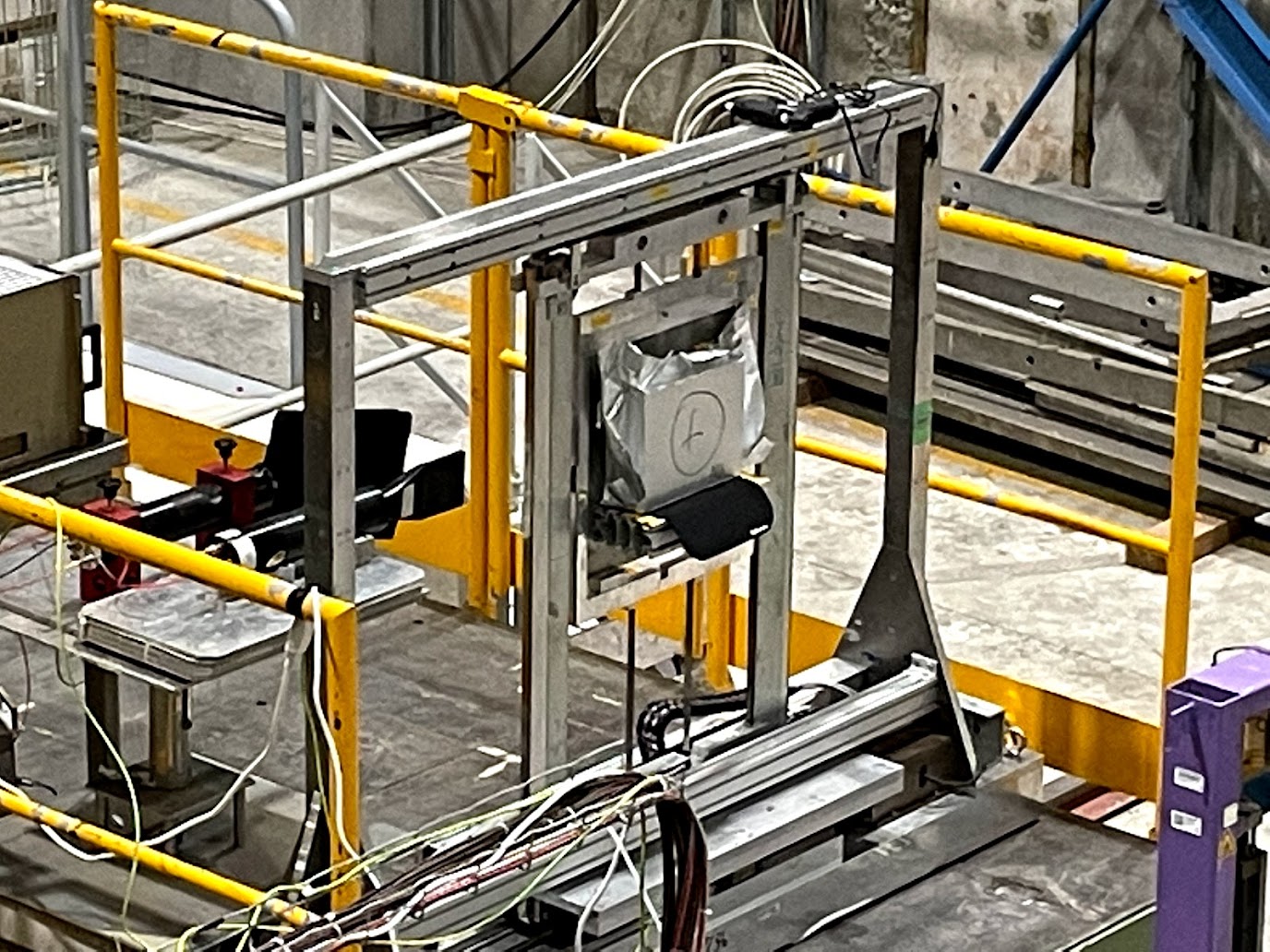}
\caption{\label{fig:b} Left: The schematic view of the Target Mover used in the J-PARC E07~\cite{J-PARC:2023}. The overall size of the structure is 1370 \si{mm} height, 1800 \si{mm} width, and 400 \si{mm} depth, and the range of the motion is 350 \si{mm} in the vertical direction and 450 \si{mm} in a horizontal one. Right: The picture of the Target Mover used in the DsTau 2021 physics run at the SPS H2 beamline with a detector module and the stage module for mounting it.}
\end{figure}

The detector modules are exposed to the CERN SPS 400 GeV proton beam with an intensity order of $10^5$ per spill with a duration of about 4 seconds. 
The emulsion accumulates the trajectory of charged particles passing through, however, there is a limit of the track density which can be successfully processed and analyzed.
As the proton beam spot is small, the target mover (TM) system (as shown in Figure~\ref{fig:b}) is utilized to uniformly irradiate the whole surface of the emulsion detectors.
The similar movable stages were used in the past experiments~\cite{Kodama:1990zy, Aoki:1988pm, KONOVALOVA2019100401}.
The small scale TM prototype was used during the test runs in 2016 and 2017 then in the pilot run of 2018~\cite{DsTau:2019wjb}.
As the detector modules used for 2021 physics run were four times larger than those used in the test and pilot runs, the payload and moving range of the TM should be >\SI{20}{kg} and >$\SI{350}{cm}\times\SI{350}{cm}$, respectively. 
Thus, a new TM with a wide aperture was developed by modifying the TM used in another emulsion experiment, J-PARC E07~\cite{J-PARC:2023}, and adding a new functionality to move the stage of TM with a speed proportional to the beam intensity specifically for the DsTau experiment. 
This paper reports on the development of the new TM and control system, and evaluates their performance in the 2021 physics run.

\section{The Target Mover and the real-time speed control system}
\begin{figure}[tbp]
\centering
\includegraphics[width=.8\textwidth]{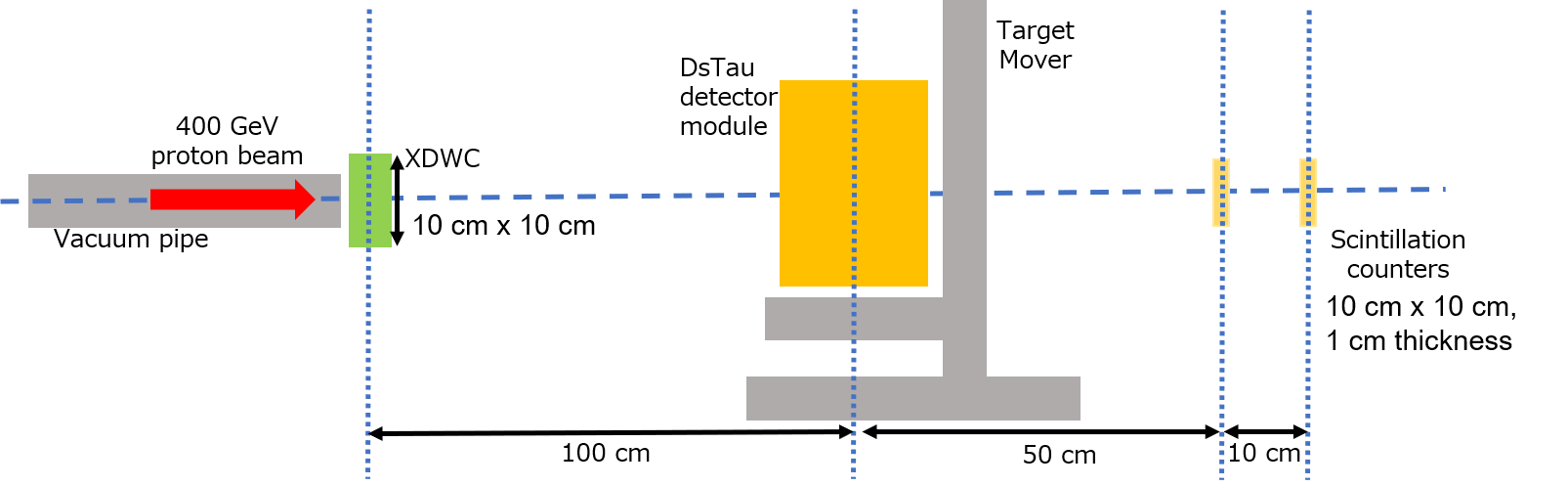}
\caption{\label{fig:setup_SPS}The Schematic view of the experimental setup of the beam test at CERN SPS.}
\end{figure}

\begin{figure}[tbp]
\centering
\includegraphics[width=.9\textwidth]{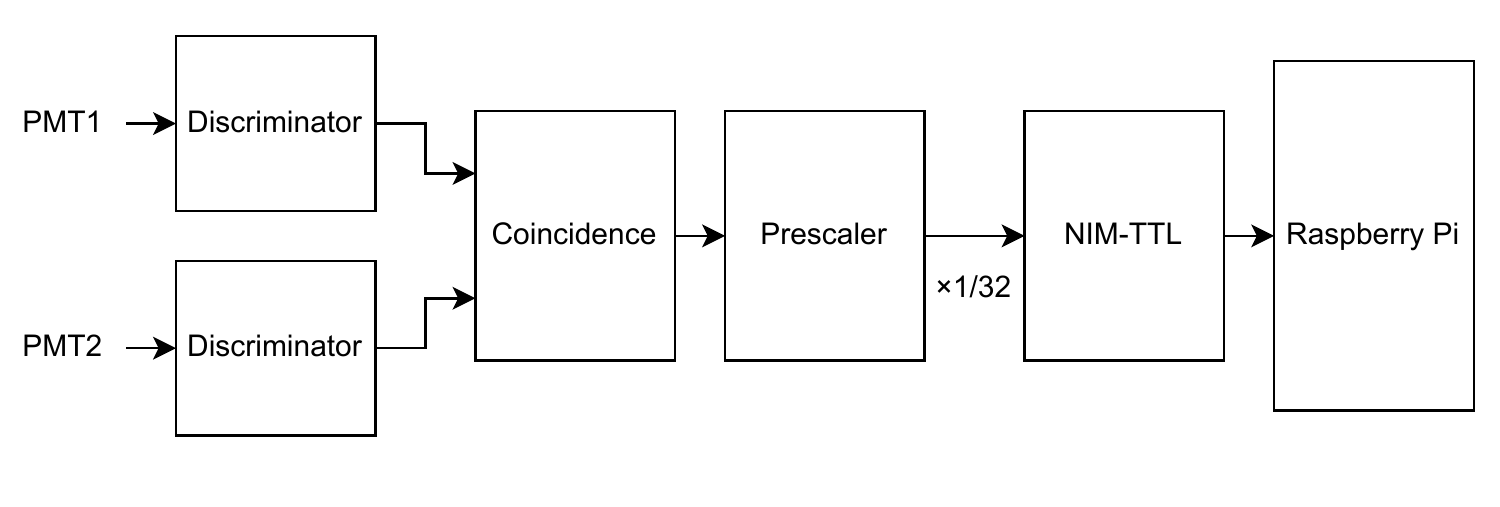}
\caption{The flowchart of the signal conversion from the scintillation counters to the Raspberry Pi. 
}
\label{fig:SigConv}
\end{figure}

\begin{figure}[tbp]
\centering
\includegraphics[width=.9\textwidth]{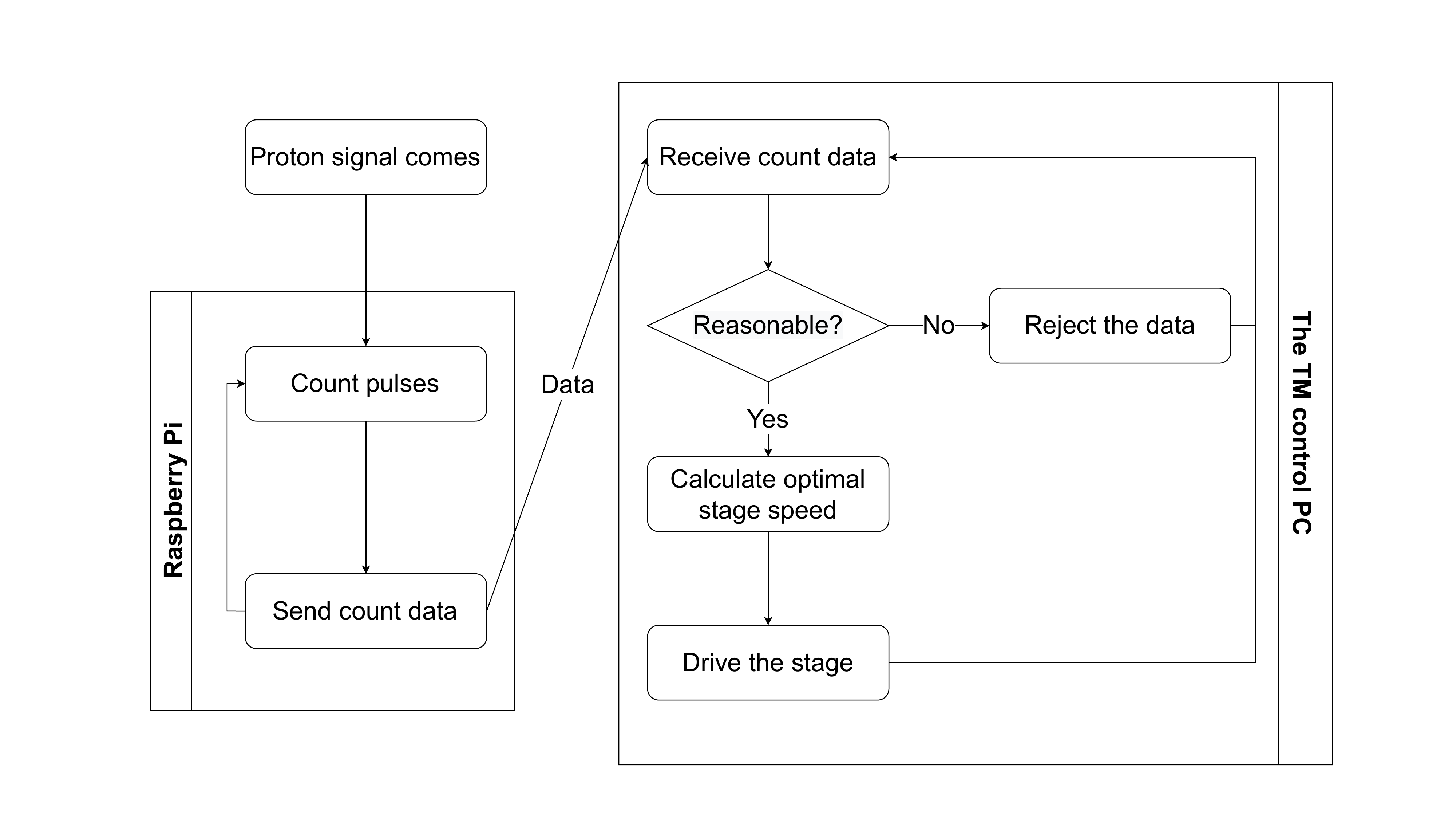}
\caption{The flowchart of the RSCS. The left side shows the proton counting process and the right shows the TM control process.
The condition "Reasonable" means that the count data is not too high or low compared with the previous sent count.
In order to prevent the excessive acceleration of the motor, the RSCS rejects counts when that is either more than 50 times or less than 1/50 of the previous count and does not change the stage speed. 
Such "Not reasonable" counts are mainly caused by bit errors in the data transmission. 
This was observed at the development stage, but not in the physics run.}
\label{fig:c}
\end{figure}

The TM is a motorized 2-dimensional stage to raster-scan the emulsion module with respect to the beam.
The stepping motors drive the stage under the control of a computer with a program written in C\# language.
Stepping motors offer more precise control over stage position and speed compared to the DC motors that were used to control the TM in previous experiments~\cite{Kodama:1990zy, Aoki:1988pm}.
We have implemented additional mechanical support to hold the emulsion modules.
The schematic view of the experimental setup is shown in Figure~\ref{fig:setup_SPS}.
The cross delayed wire chamber (XDWC) measures the proton beam profile.
The hit efficiency of XDWC we used was too low, $<20\%$, to be used as the proton counter.
Therefore, two scintillation counters to obtain proton counts were located behind the TM.
Coincidence were taken in order to minimize the contamination from backgrounds.
The trigger threshold of them was set to well below the MIP level.
As shown in Figure~\ref{fig:SigConv}, signals from the scintillation counters are sent to a series of NIM modules (discriminator, coincidence module, pre-scaler, and NIM-TTL converter). 
The coincidence signal is then transferred to a Raspberry Pi 4B microcomputer.
The Raspberry Pi counts the pulses and sends the data to the TM control PC every 100 \si{ms}.
A TCP-IP protocol is used for the communication between the Raspberry Pi and the TM control PC.
The TM control PC calculates the optimal stage speed $v_x$ based on the following formulas:
\begin{equation}
\label{eq:y}
\begin{split}
v_x = \frac{I}{\rho\Delta y} \,,
\qquad
I \equiv \frac{\Delta n}{\Delta t} \, , 
\end{split}
\end{equation}
where $\Delta n$ is the count taken by the Raspberry Pi, $\Delta t$ is the time interval of their count measurement ($\sim$ 100 \si{ms}), $\rho$ is the required proton density ($\sim 10^5$ \si{cm^{-2}} for physics run) and $\Delta y$ is the $y$-step size of the raster-scanning, which depends on the beam profile as discussed in Section \ref{sec:2021}.
Figure~\ref{fig:c} shows the flowchart of this system which is called the real-time speed control system (RSCS).


At the beginning, the stage moves to the start position. 
When the beam is exposed on the module, the stage moves in $x$ direction for 330 \si{mm} at the speed controlled by the RSCS, then moves in $y$ direction by $y$-step at a constant speed of $\SI{5}{mm/s}$.
Then the stage moves again along $x$ axis, but in the opposite direction.
The TM repeats these steps to expose the entire surface of the module. 
We avoided scanning in y direction because the detector module is so heavy and may affect the RSCS performance. 
After the module irradiation is completed, the stage goes back to the start position.
In the event of any trouble, the operator could immediately stop the TM and return it to the start position where the detector would not be exposed to the beam. 
Once the issue is resolved, the raster scan could be restarted from where it is interrupted.

The requirement on the proton density is \SI{e5}{cm^{-2}} with less than 10\% fluctuation. 
This would cause 1\% systematic uncertainty in $D^0$ detection due to fluctuation of misidentified $K^0$.
To be precise, we seeking for $D^0$ by searching for neutral decay around proton interaction vertex. 
Since emulsions does not have time information, neutral decay (e.g. $K^0$) from other proton interaction can be in the region of interest, being the background to $D^0$. 
In order to estimate the background ratio, we will compare the data with the MC simulation with the uniform proton density. 
This is the level of uncertainties that we can tolerate for this analysis.

\section{Testing and commissioning of the TM}
\begin{figure}[tbp]
\centering
\includegraphics[width=.4\textwidth]{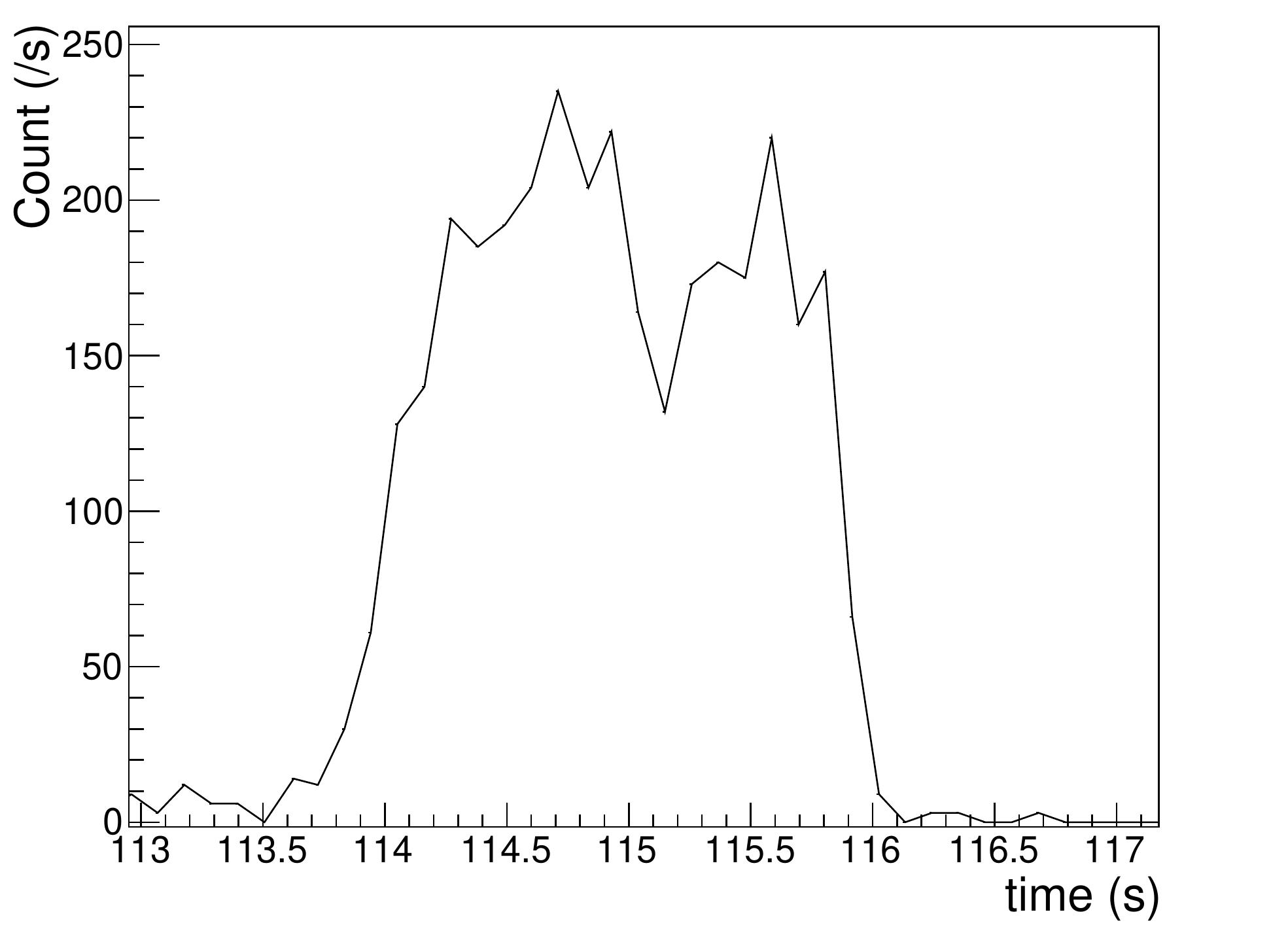}
\caption{\label{fig:e}The emulated beam spill by the \si{^{90}Sr} source. The counts are plotted every \SI{100}{ms} and they are normalized to the counts per second.}
\end{figure}

The RSCS performance was evaluated 
using \si{^{90}Sr} $\beta$ source in Chiba University. 
A lead brick with a weight of 22.7 kg was mounted on the TM instead of the emulsion module, and a 2 \si{kBq} \si{^{90}Sr} source was used to emulate the SPS beam. 
Since electrons from the source did not penetrate through both scintillators, the count rate was too low.
Therefore, we employed only one scintillation counter and omit the prescaler. 
Figure~\ref{fig:e} shows an example of the emulated beam spills using the \si{^{90}Sr} source. 
The pseudo beam spills were emulated by placing the source on the scintillation counter for 2.5 seconds by hand, this was repeated about 950 times.

The speed of the TM stage was optimized using data of the 2018 pilot run at the SPS and calculated by Equation~\ref{eq:y}.
In order to emulate the stage velocity similar to one of the real experiments, the required pseudo beam density
and $y$-step size of raster-scanning were set to \SI{1500}{cm^{-2}} and \SI{12}{mm}, respectively. 

\begin{figure}[tbp]
\centering
\includegraphics[width=.4\textwidth]{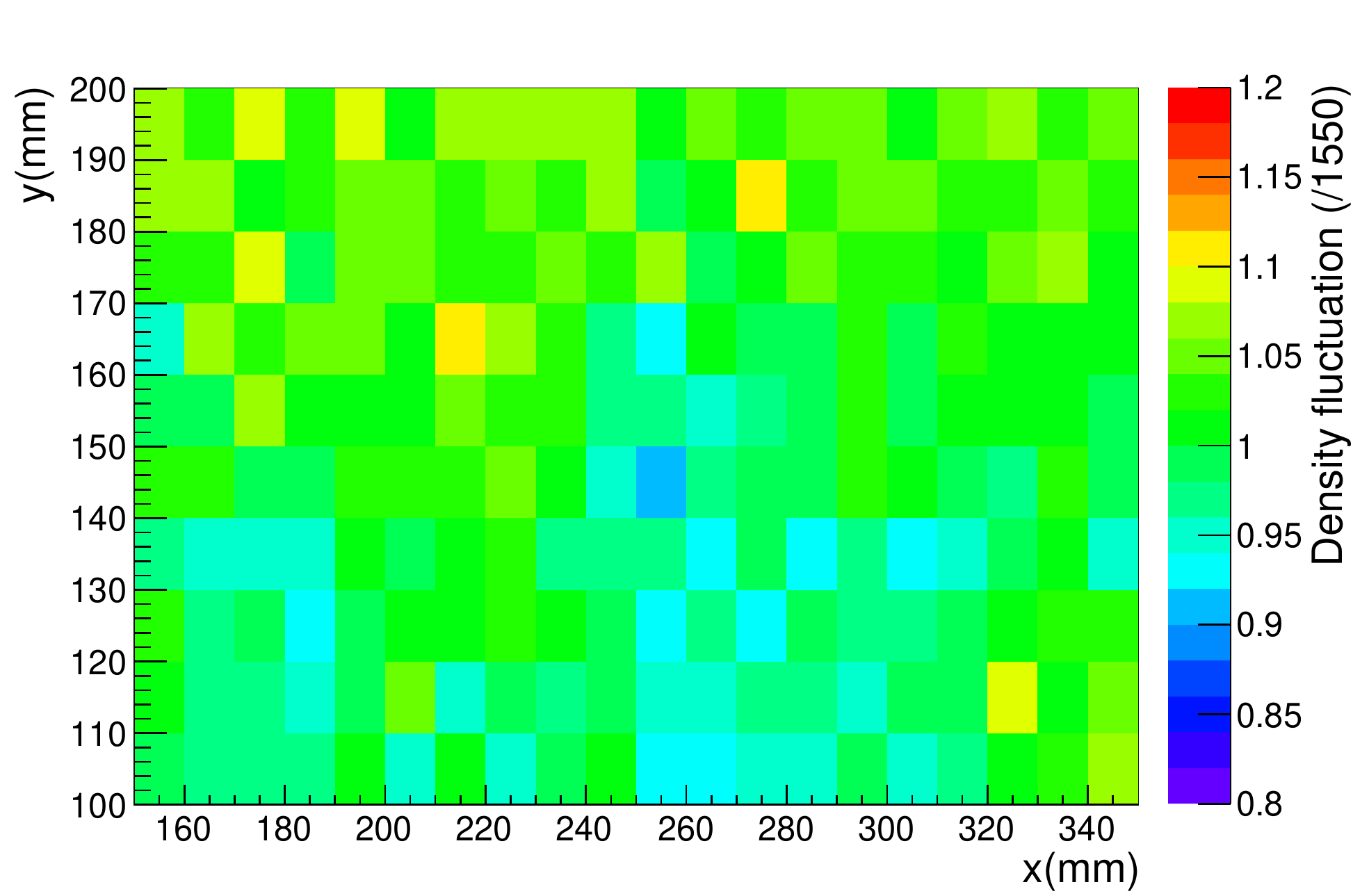}
\qquad
\includegraphics[width=.4\textwidth]{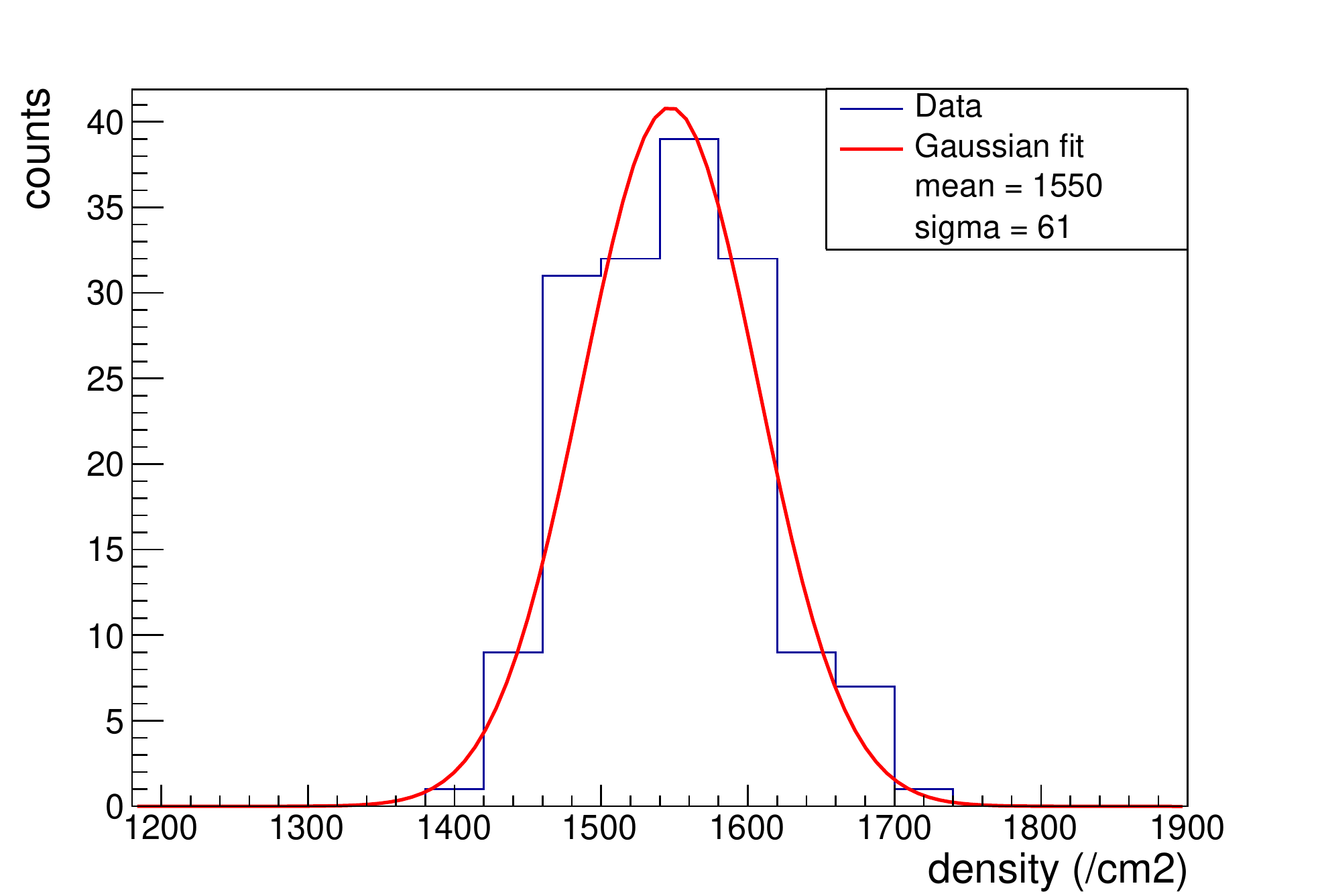}
\caption{\label{fig:g}Left: The map of the emulated particle density fluctuation map of the TM test. Right: 
The emulated particle density distribution with Gaussian fit.}
\end{figure}

To emulate the proton density map, the recorded stage position and source intensity were smeared by weighting each signal by a 2D Gaussian, with a sigma of \SI{10}{mm}.
Figure~\ref{fig:g} shows the calculated density map and distribution of the density in each 1 \si{cm^2}.
The density was measured as $\SI{1550 +- 61}{cm^{-2}}$.
The achieved uniformity is approximately $4\si{\%}$, which satisfies the requirement of the experiment ($< 10\si{\%}$). 

\section{Physics run at the CERN SPS H2 beamline}
\label{sec:2021}
The NA65/DsTau 2021 physics run was performed at the CERN SPS H2 beamline from $23^\textnormal{rd}$ of September to $7^\textnormal{th}$ of October in 2021.

\begin{figure}[tbp]
\centering
\includegraphics[width=.6\textwidth]{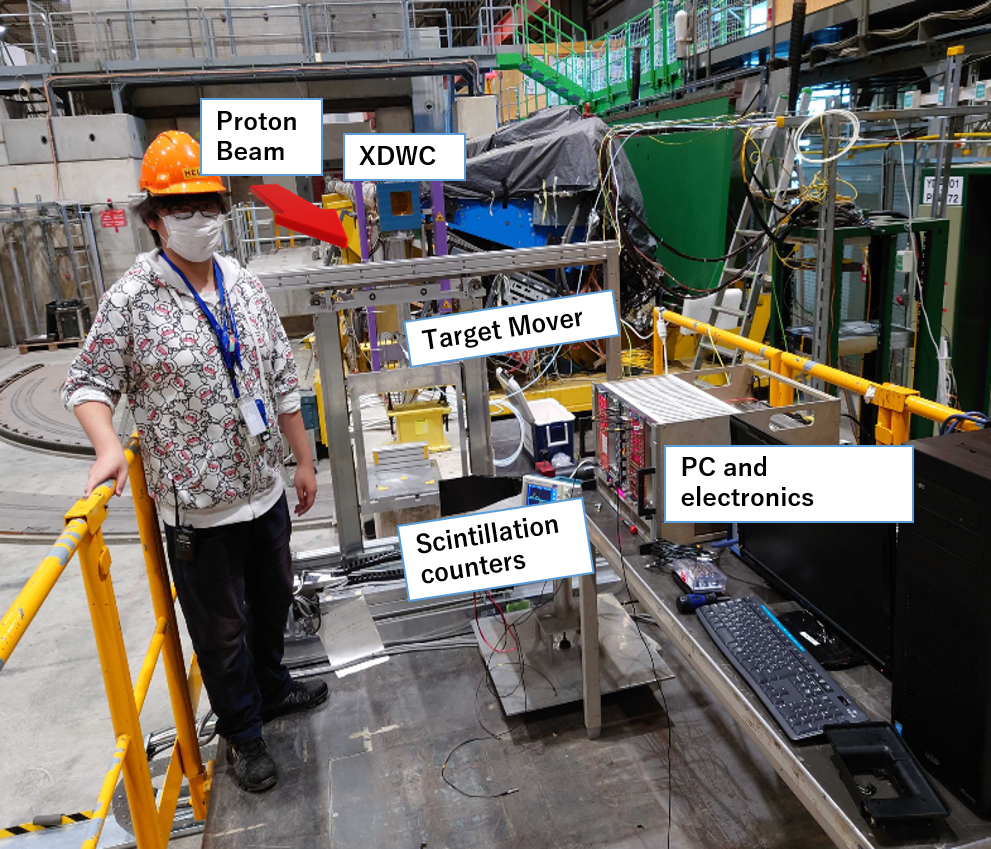}
\caption{\label{fig:TM_SPS}
The experimental setup placed at the CERN SPS H2 beamline for the NA65/DsTau 2021 physics run.}
\end{figure}

\begin{figure}[tbp]
\centering
\includegraphics[width=.5\textwidth]{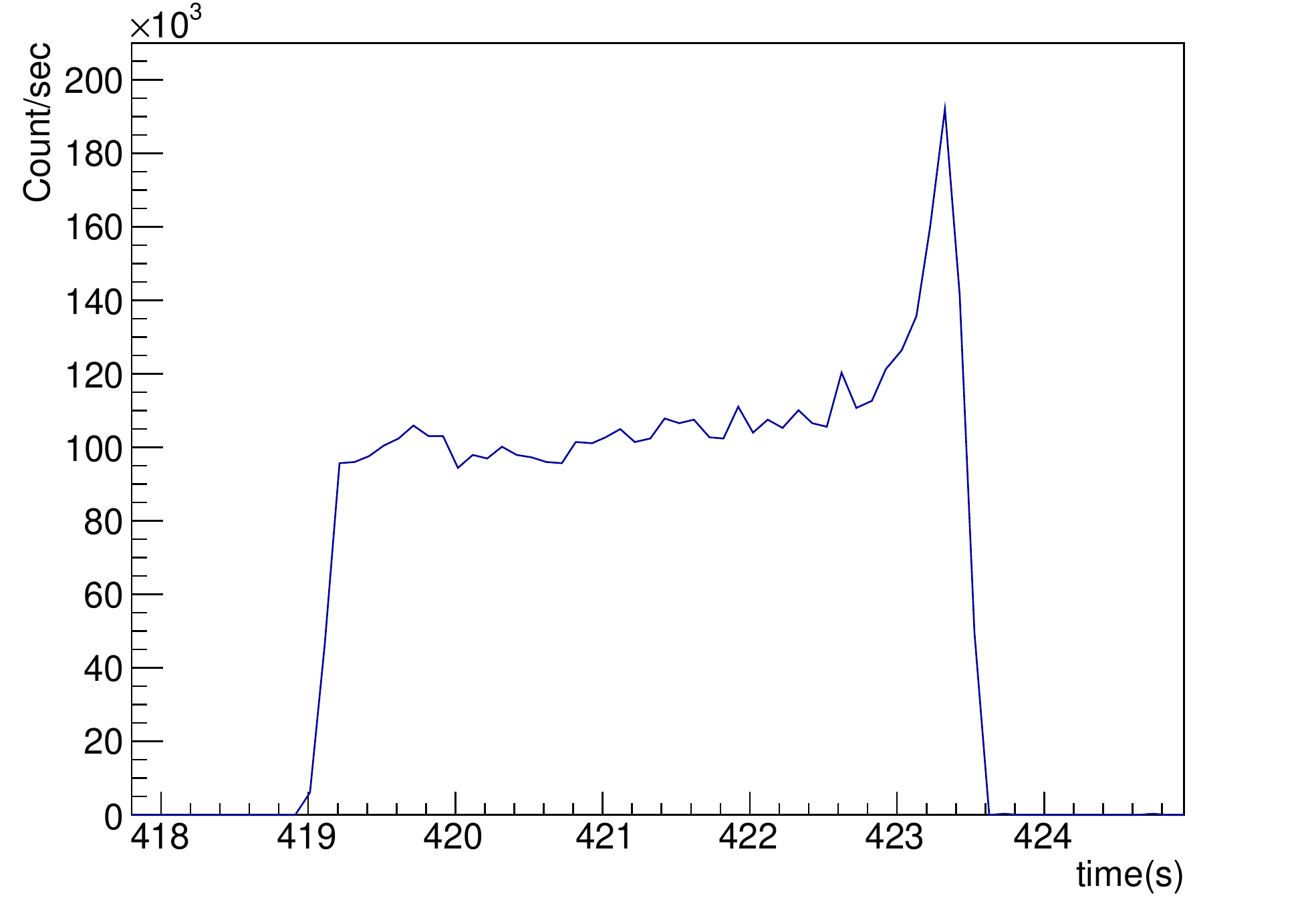}
\caption{\label{fig:beam_intensity_SPS}
 Example of the time profile of the beam. In certain spills, we observed that  the extracted beam from the accelerator was not very uniform in the time domain, due to quadruple ripples or other effects that from time-to-time affect the slow extraction towards the North Area and the H2 beamline. However, this was compensated and was not problematic for the data taking.}
\end{figure}

\begin{figure}[tbp]
\centering
\includegraphics[width=.4\textwidth]{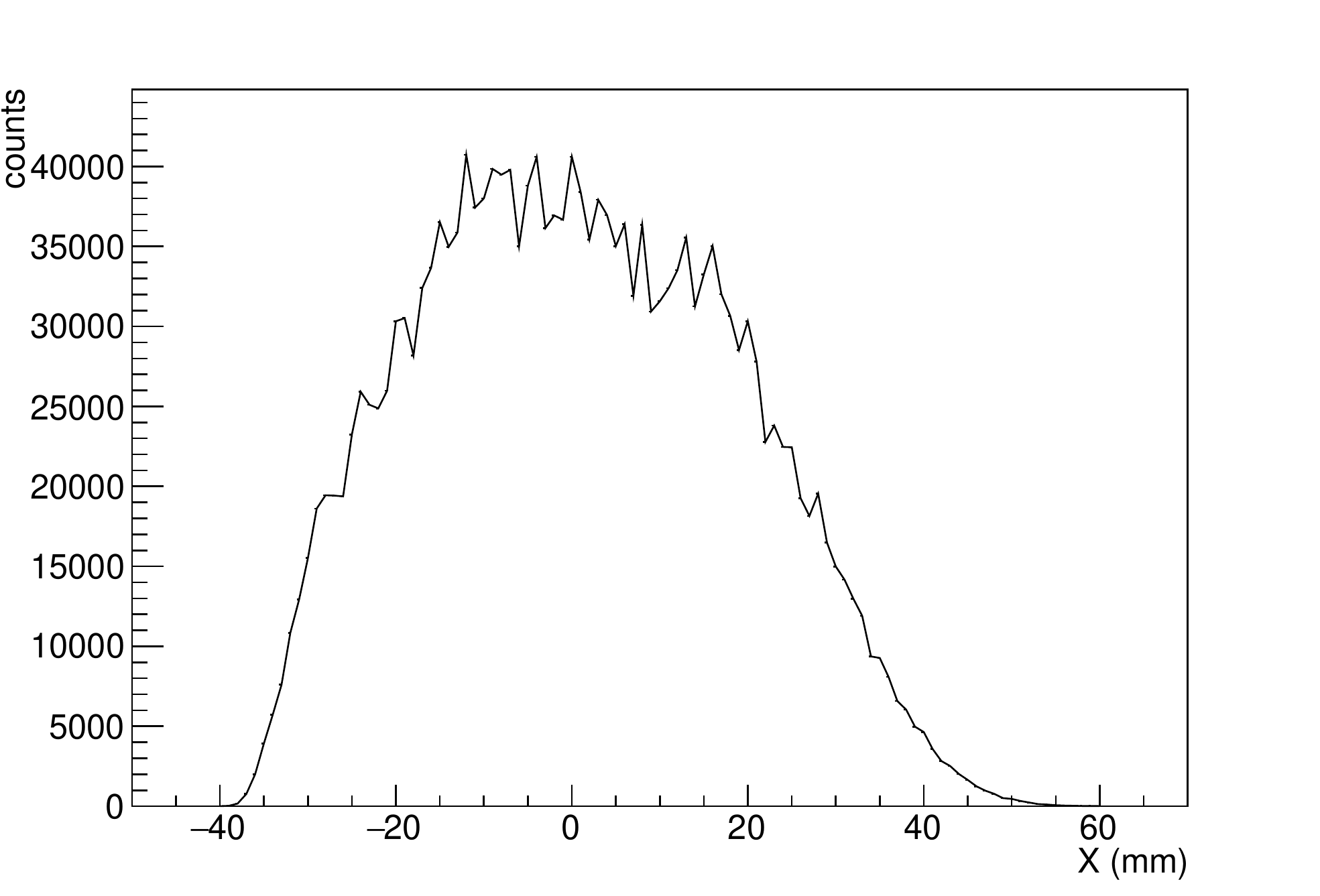}
\qquad
\includegraphics[width=.4\textwidth]{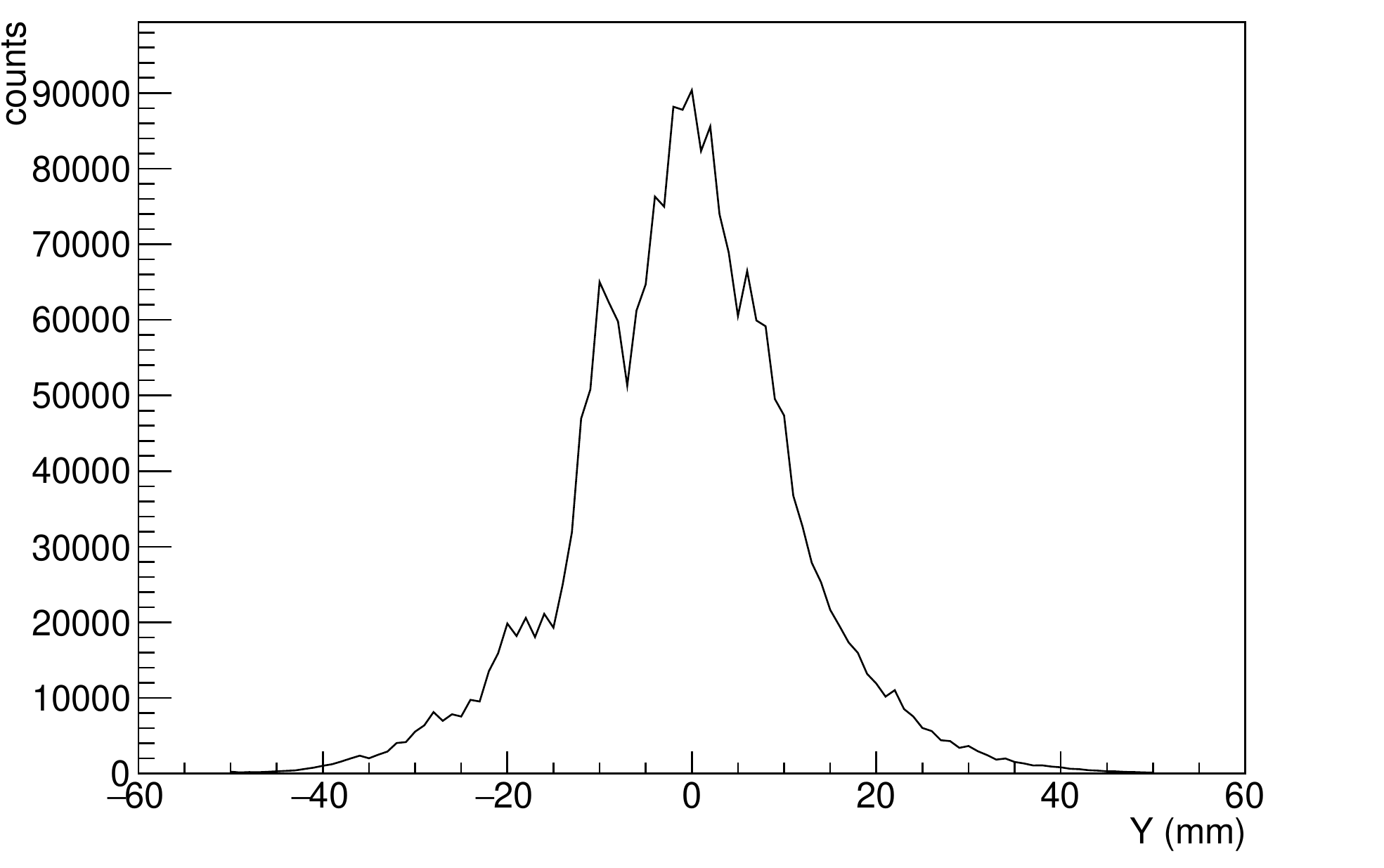}
\caption{\label{fig:h}Left: X (horizontal) profile of the NA65 beam in 2021. Right: Y (vertical) profile of the NA65 beam in 2021.}
\end{figure}

\begin{figure}[tbp]
\centering
\includegraphics[width=1.0\textwidth]{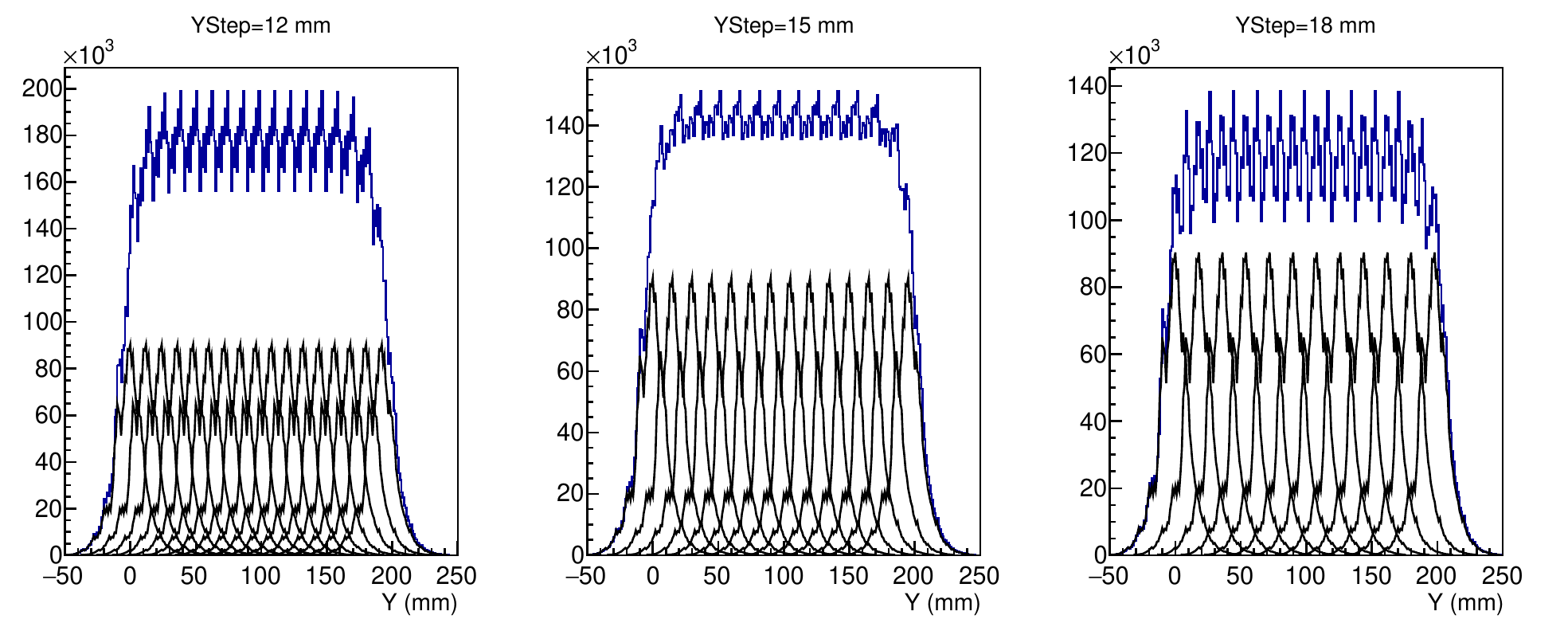}
\caption{\label{fig:ySteps}
The overall protons density distribution as a superposition of several beam profiles with different $y$-steps of \SI{12}{mm}, \SI{15}{mm}, and \SI{18}{mm}. The deviation is larger in $y$-step \SI{12}{mm} case than \SI{15}{mm} case due to the Y (vertical) profile of the SPS beam having 2nd peaks at around $\pm \SI{12}{mm}$, as Figure~\ref{fig:h} shows.}
\end{figure}

As shown in Figure~\ref{fig:TM_SPS}, all components were implemented at the CERN SPS H2 beamline. 
The Raspberry Pi of the RSCS was operated from the control room, outside of the beam area. 
The count data from the Raspberry Pi was transmitted to the TM control PC via Ethernet. 
The ping value of their communication was less than $\SI{1}{msec}$, the minimum value detectable by the Windows OS.
This network delay is significantly shorter than \SI{100}{ms}, tha data transfer cycle, and negligible for the RSCS performance. 
The beam energy was \SI{400}{GeV}, with an intensity of $5 \times 10^5$ particles per spill of about 4 seconds.
The beam intensity had a time structure as shown in Figure~\ref{fig:beam_intensity_SPS}, and as the peak shows that the data taking rate was up to about \SI{200}{kHz}.
The beam profile measured by the XDWC 
demonstrated RMS values of $\SI{12}{mm}$ in $x$ direction and $\SI{13}{mm}$ in $y$. 
We assumed that the position profile would not change during the irradiation. 
And the operators periodically monitored the beam profiles and the motion of the TM, and no significant deviation of profiles was observed, except for occasional accelerator failures. 
Figure~\ref{fig:h} shows a measured beam profile in 
$y$ (vertical) direction. 
In order to find an optimal $y$-step value, the profile was multiple-copied and superimposed with different $y$-steps as shown in Figure~\ref{fig:ySteps}. 
The step size was determined to be \SI{15}{mm} to flatten the 
overall proton density distribution.

\begin{figure}[tbp]
\centering  \includegraphics[width=.6\textwidth]{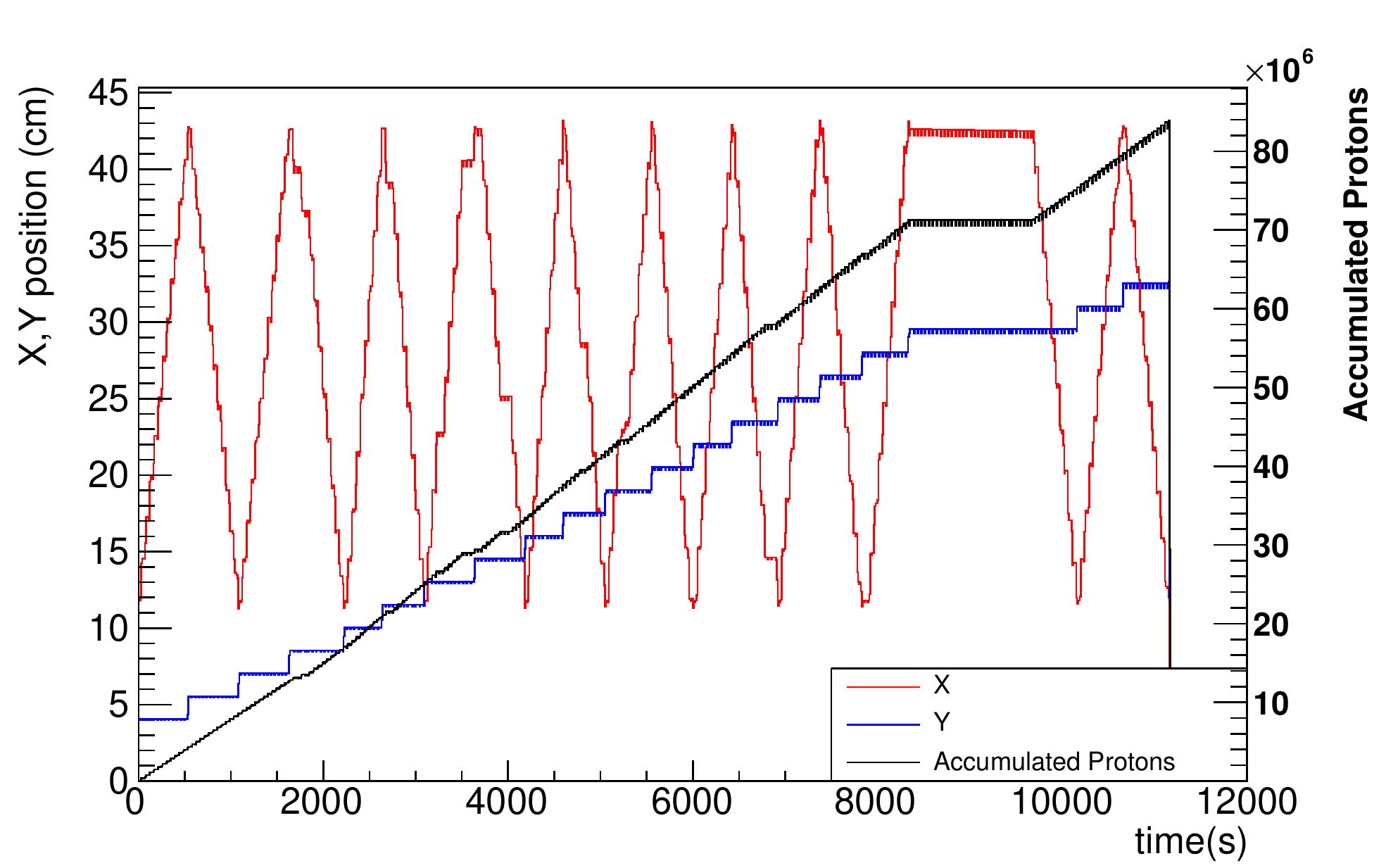}    \caption{The diagram of the TM position and the accumulated number of protons as a function of time for one of the modules in the 2021 physics run. The red line shows the 
$x$ position going right and left. 
The blue line shows the $y$ position going up step by step. 
The black line shows the 
integrated number of protons that have passed the emulsion module. 
The flat part corresponds to the period without beam. }
\label{fig:raster}
\end{figure}

The TM was successfully operated during the 2021 run for 17 detector modules with the size of $\SI{25}{cm} \times \SI{20}{cm} \times \SI{7}{cm}$. 
Figure~\ref{fig:raster} describes the TM sequence for one of the modules in the 2021 run.
Together with the data collected in the 2018 pilot run, approximately 30\% of the total amount of the proton interactions, planned to be registered in the experiment, were accumulated.

\begin{figure}[tbp]
\centering
\includegraphics[width=.4\textwidth]{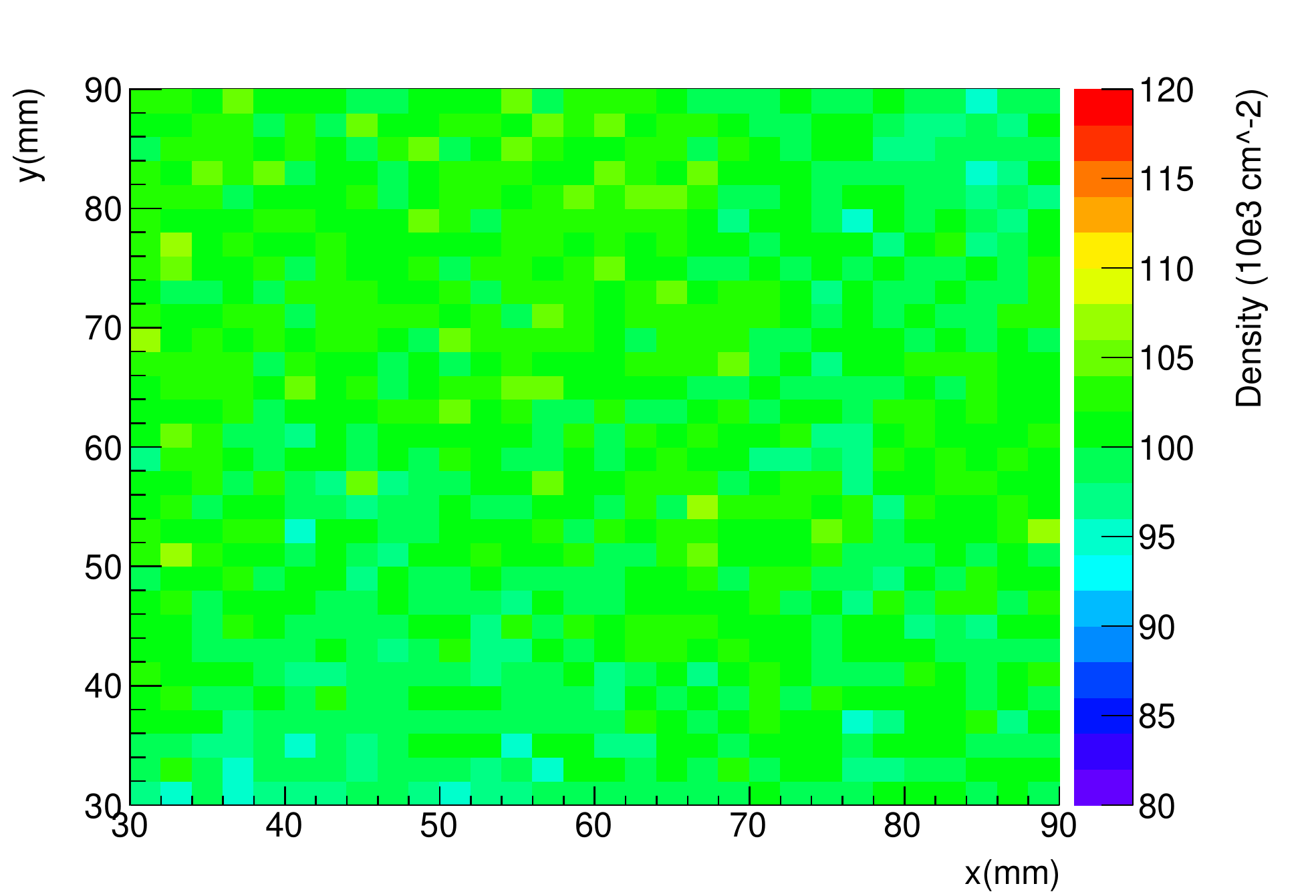}
\qquad
\includegraphics[width=.4\textwidth]{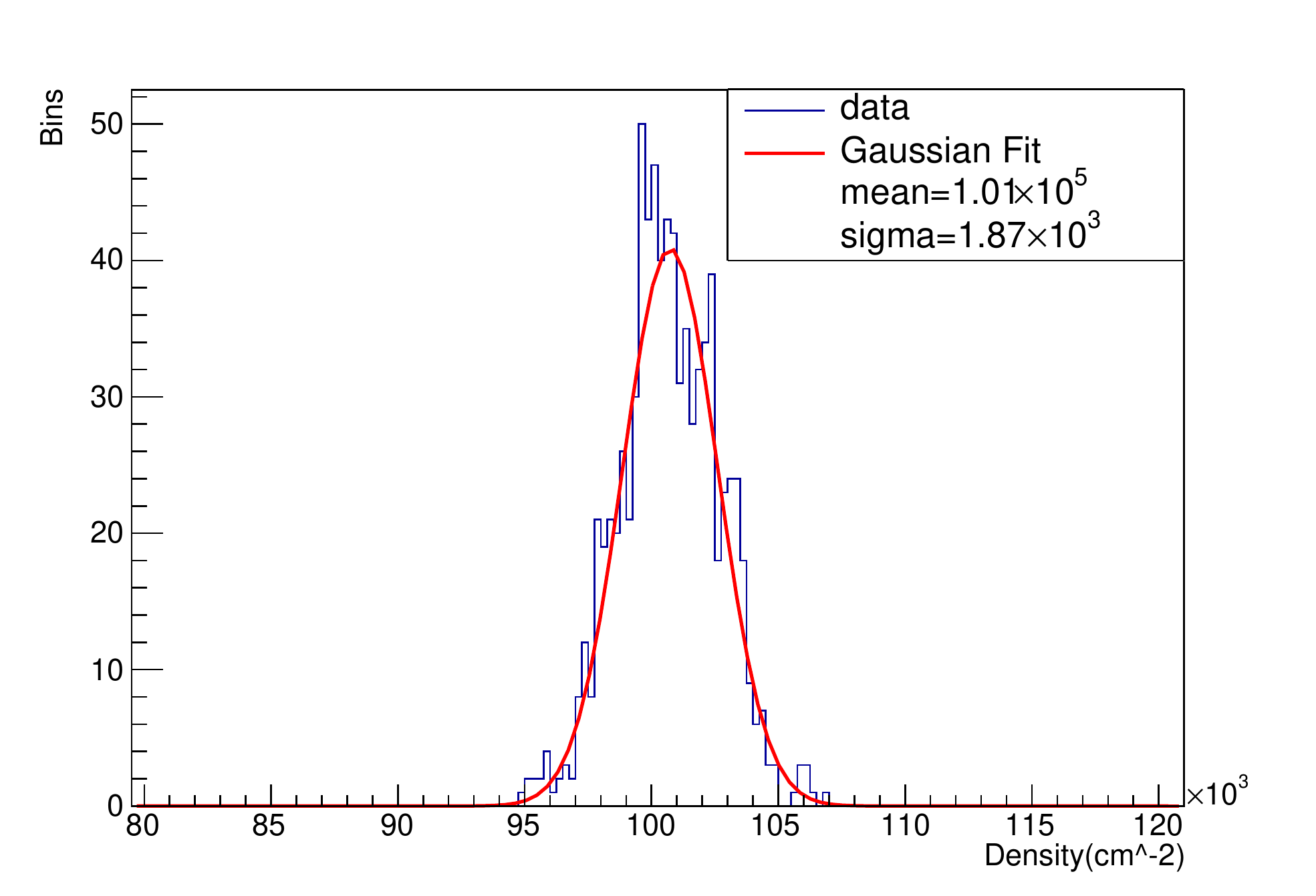}
\caption{\label{fig:i}Left: Track density fluctuation mapping on the emulsion of the DsTau 2021 physics run. Right: The density distribution histogram with Gaussian fitting.}
\end{figure}

After the data taking in H2, the irradiated emulsion modules were dismantled, then the films were developed chemically at the CERN nuclear emulsion facility.
These developed films were transported to Nagoya University 
and scanned by the HTS \cite{10.1093/ptep/ptx131}, the high-speed automatic microscope.
To analyze the density of registered primary protons, the tracks were reconstructed in the 
10 most upstream emulsion films. 
Figure~\ref{fig:i} shows the measured proton track density map with the bin size of $\SI{2}{mm} \times \SI{2}{mm}$, normalized to \SI{1}{cm^2}, and distribution of the proton density in this data sample.


The mean value of the density is $\mu_\rho = \SI{1.01e5}{cm^{-2}}$, and the standard deviation of the fitted Gaussian function is $\sigma_\rho = 0.0187 \times 10^5$~\si{cm^{-2}}.
The fluctuation of the proton density is $\sigma_\rho/\mu_\rho = 1.9\pm0.3\%$, which satisfies the requirement of $< 10\%$ fluctuation.

\section{Summary}
The NA65/DsTau experiment aims to study tau neutrino production by detecting $D_s \to \tau \to X$ events with the emulsion based detector.
The proton distribution on the detector surface should be uniform with the density of $10^5$ \si{cm^{-2}}.
For physics analysis, the proton density fluctuation should be $<10\%$ and the data taking rate should be $\mathcal{O}(10^5)$ Hz.
With the help of the new TM and the RSCS, fluctuation $<10\%$ in the pseudo proton density was achieved in the commissioning run with the radioactive source.
During the DsTau physics run in 2021 at the CERN-SPS H2 beamline, the TM and the RSCS worked successfully and allowed the data taking rate of \SI{200}{\kilo\hertz} and the proton density of $\SI{1.01e05}{cm^{-2}}$ with $ 1.9\pm0.3\%$ fluctuation, which exceeds the requirement of the DsTau experiment. 

\acknowledgments

We thank for the members of the J-PARC E07 experiment for allowing us to use the TM mechanics.

Funding is gratefully acknowledged from national agencies and Institutions supporting us,
namely:   JSPS KAKENHI for Japan (Grant No. JP 20K23373, JP 18KK0085, JP 17H06926, JP 18H05541), 
CERN-RO(CERN Research Programme)   for Romania (Contract No. 03/03.01.2022)  and  TENMAK for Turkey (Grant No. 2022TENMAK(CERN) A5.H3.F2-1).

\bibliography{main}

\end{document}